\documentclass[12pt]{iopart} 
\usepackage{epsfig,amssymb}  
\begin{document} 
\title[]{Determining the WIMP mass from a single 
direct detection experiment,
a more detailed study} 
\author{Anne M.~Green\dag}
\address{\dag\ School of
Physics and Astronomy, University of Nottingham, University Park,
  Nottingham, NG7 2RD, UK}
\eads{\mailto{anne.green@nottingham.ac.uk}}, 
  \begin{abstract} 
  The energy spectrum of nuclear recoils in Weakly Interacting Massive
  Particle (WIMP) direct detection experiments depends on the
  underlying WIMP mass.  We study how the accuracy with which the WIMP
  mass could be determined by a single direct detection experiment
  depends on the detector configuration and the WIMP properties. We
  investigate the effects of varying the underlying WIMP mass and
  cross-section, the detector target nucleus, exposure, energy
  threshold and maximum energy, the local circular speed and the
  background event rate and spectrum.  The number of events observed
  is directly proportional to both the exposure and the cross-section,
  therefore these quantities have the greatest bearing on the accuracy
  of the WIMP mass determination.  The relative capabilities of
  different detectors to determine the WIMP mass depend not only on
  the WIMP and target masses, but also on their energy thresholds. The
  WIMP and target mass dependence of the characteristic energy scale
  of the recoil spectrum suggests that heavy targets will be able to
  measure the mass of a heavy WIMP more accurately. We find, however,
  that the rapid decrease of the nuclear form factor with increasing
  momentum transfer which occurs for heavy nuclei means that this is
  in fact not the case.  Uncertainty in the local circular speed and
  non-negligible background would both lead to systematic errors in
  the WIMP mass determination.  For deviations of $\pm 20 \, {\rm km
    \, s}^{-1}$ in the underlying value of the circular speed the
  systematic error is of order $10\%$, increasing with increasing WIMP
  mass.  This error can be reduced by also fitting for the circular
  speed.  With a single detector it will be difficult to disentangle a
  WIMP signal (and the WIMP mass) from background if the background
  spectrum has a similar shape to the WIMP spectrum (i.e. exponential
  background, or flat background and a heavy WIMP).

\end{abstract}

 
\begin{flushleft} {\bf Keywords}: dark matter, dark matter detectors
\end{flushleft} 
 
 
\maketitle 

\vspace{0.3cm}

\section{Introduction}

Cosmological observations indicate that the majority of the matter in
the Universe is dark and non-baryonic (e.g. Ref.~\cite{cosmo}).
Weakly Interacting Massive Particles (WIMPs) are one of the leading
cold dark matter candidates, and supersymmetry provides a concrete,
well-motivated WIMP candidate in the form of the lightest neutralino
(e.g. Ref.~\cite{jkg,dmrev}).  The direct detection of WIMPs in the
lab~\cite{ddtheory} (see Ref.~\cite{ddexpt} for a review of current
and future experiments) would not only directly confirm the existence
of dark matter but would also allow us to probe the WIMP properties, as
the shape of the differential event rate depends on the WIMP
mass~\cite{ls,lf,bk,brown,pap1,sd}. Constraints on, or measurements
of, the WIMP mass and elastic scattering cross-section will be complementary
to the information derived from collider and indirect detection
experiments~\cite{collider}.

In paper I~\cite{pap1}, see also Refs.~\cite{brown,sd}, we examined the
accuracy with which a future SuperCDMS~\cite{SuperCDMS} like direct
detection experiment would be able to measure the WIMP mass, given a
positive detection. In this paper we revisit that analysis, studying
in more detail the dependence of WIMP mass limits on the detector
capabilities (including threshold energy, exposure, maximum energy and
target nucleus), the WIMP properties (mass and cross-section), the
local circular speed and the effects of non-zero backgrounds. We
outline the calculation of the differential event rate and the
Monte Carlo simulations in Sec.~\ref{method} (see paper I~\cite{pap1} for
further details), present the results in Sec.~\ref{results} and
conclude with discussion in Sec.~\ref{discuss}.

\section{Method}
\label{method}
\subsection{Event rate calculation}
\label{secdrde}
The differential event rate,  assuming 
spin-independent coupling, 
is given by (see e.g.~\cite{jkg,ls}):
\begin{equation}
\label{drde}
\frac{{\rm d} R}{{\rm d}E}(E) =
             \frac{\sigma_{{\rm p}} 
             \rho_{\chi}}{2 \mu_{{\rm p} \chi}^2 m_{\chi}}
             A^2 F^2(E) {\cal F}(E)   \,, 
\end{equation}
where $\rho_{\chi}$ is the local WIMP density, $\sigma_{{\rm
p}}$ the WIMP scattering cross section on the proton, $\mu_{{\rm p} \chi} = 
(m_{\rm p} m_{\chi})/(m_{{\rm p}}+ m_{{\chi}})$ the WIMP-proton reduced mass, 
$A$ and $F(E)$ the mass number and
form factor of the target nucleus respectively and $E$ is the recoil
energy of the detector nucleus. We use the Helm form factor~\cite{helm}.
The dependence  on the WIMP velocity distribution is encoded in
${\cal F}(E)$, which is defined as
\begin{equation}
\label{tq}
{\cal F}(E)= \langle \int^{\infty}_{v_{{\rm min}}} 
            \frac{f^{\rm E}(v,t)}{v} {\rm d}v  \rangle \,,
\end{equation}
where $f^{\rm E}(v,t)$ is the (time dependent) 
WIMP speed distribution in the rest frame of the
detector, normalized to unity and $\langle .. \rangle$ 
denotes time averaging. The WIMP speed distribution
is calculated from the
velocity distribution in the rest frame of the Galaxy,
$f^{\rm G}({\bf v})$, via Galilean transformation:
${\bf v} \rightarrow \tilde{{\bf v}}= {\bf v} + {\bf v}^{\rm E}(t)$
where ${\bf v}^{\rm E}(t)$ is the Earth's velocity with
respect to the Galactic rest frame~\cite{ls}.
The lower limit of the integral, $v_{{\rm min}}$, is the minimum
WIMP speed that can cause a recoil of energy $E$:
\begin{equation}
\label{vmin}
v_{{\rm min}}=\left( \frac{ E m_{A}}{2 \mu_{{\rm A} \chi}^2} 
             \right)^{1/2} \,,
\end{equation}
where $m_{A}$ is the atomic mass of the detector nuclei
and $\mu_{{\rm A} \chi}$ the WIMP-nucleon reduced mass.
We use  the `standard halo model',
an isotropic isothermal sphere, for which
 the local WIMP velocity distribution, in the Galactic
rest frame, is Maxwellian 
\begin{eqnarray}
f^{\rm G}({\bf v}) & = & N \left[ \exp{\left(- 
    |{\bf v}|^2/v_{\rm c}^2 \right)} -
         \exp{\left(- v_{\rm esc}^2/v_{\rm c}^2 \right)} \right] 
     \hspace{1.0cm} |{\bf v}|< v_{\rm esc} \,, \\
f^{\rm G}({\bf v}) & = & 0  \hspace{7.0cm} |{\bf v}|> v_{\rm esc} \,,
\end{eqnarray}
where $N$ is a normalization factor and $v_{\rm c}=220 \pm 20 \, {\rm
  km \, s}^{-1}$~\cite{klb} and $v_{\rm esc} \approx 540 \, {\rm km \,
  s}^{-1}$~\cite{rave} are the local circular and escape speeds
respectively. If the ultra-local WIMP distribution is smooth, then the
uncertainties in the detailed shape of the local velocity distribution
lead to relatively small changes in the shape
of the differential event rate~\cite{drdens,lgegreen}. Consequently
there is a relatively small, [${\cal O} (5 \%)$], systematic
uncertainty in the WIMP mass~\cite{pap1}.  We caution that the
assumption of a smooth ultra-local WIMP distribution may, 
however, not be valid on
the sub milli-pc scales probed by direct detection experiments
(e.g. Ref.~\cite{millino} but see also Ref.~\cite{milliyes} for
arguments that the ultra-local WIMP distribution consists of a large
number of streams, and is hence effectively smooth).

As shown by Lewin and Smith~\cite{ls}, see also Paper I~\cite{pap1},
 the differential event rate can,
to a reasonable approximation, be written as
\begin{equation}
\label{drde0}
\frac{{\rm d} R}{{\rm d}E}(E) = \left(\frac{{\rm d} R}{{\rm d}E}\right)_{0}
              \exp{ \left( -\frac{E}{E_{\rm R}} \right)}
              F^2(E)  \,. 
\end{equation}
The event
rate in the $E \rightarrow 0\, {\rm keV
}$ limit, $({\rm d} R/{\rm d}E)_{0}$,  and $E_{\rm R}$, the characteristic
energy scale, are given by
\begin{equation}
\label{pre}
\left(\frac{{\rm d} R}{{\rm d}E}\right)_{0}= c_{0} \frac{\sigma_{\rm p} \rho_{\chi}}
         {\sqrt{\pi} \mu_{{\rm p} \chi} ^2 m_{\chi} v_{c}} A^2 \,,
\end{equation}
and
\begin{equation}
\label{er}
E_{\rm R} = c_{E_{\rm R}} \frac{2 \mu_{{\rm A} \chi}^2 v_{c}^2}{m_{\rm A}} \,,
\end{equation}
respectively, where $c_{0}$ and $c_{E_{\rm R}}$ are constants of
order unity which are  required when the Earth's velocity and the 
Galactic escape speed are taken into account and are determined
by fitting to the energy spectrum calculated using the full expression,
eq.~(\ref{drde}).
The exact values of these constants
depend on the target nucleus, the energy threshold and the Galactic
escape speed. For a ${\rm Ge}$ detector
with energy threshold $E_{\rm th}=0 \, {\rm keV}$, $c_{0} \approx 0.78$
and $c_{E_{\rm R}} \approx 1.72$, with a weak dependence 
on the WIMP mass~\cite{pap1}. For the majority of our calculations
we will use the accurate expression, eq.~(\ref{drde}), however in 
Sec.~\ref{vcsec}
where we consider varying $v_{\rm c}$ we will use the fitting 
function, eq.~(\ref{drde0}), as it is not computationally feasible to carry out
the full calculation in the likelihood analysis in this case. 

\begin{figure}  
\begin{center}  
\epsfxsize=6.in  
\epsfbox{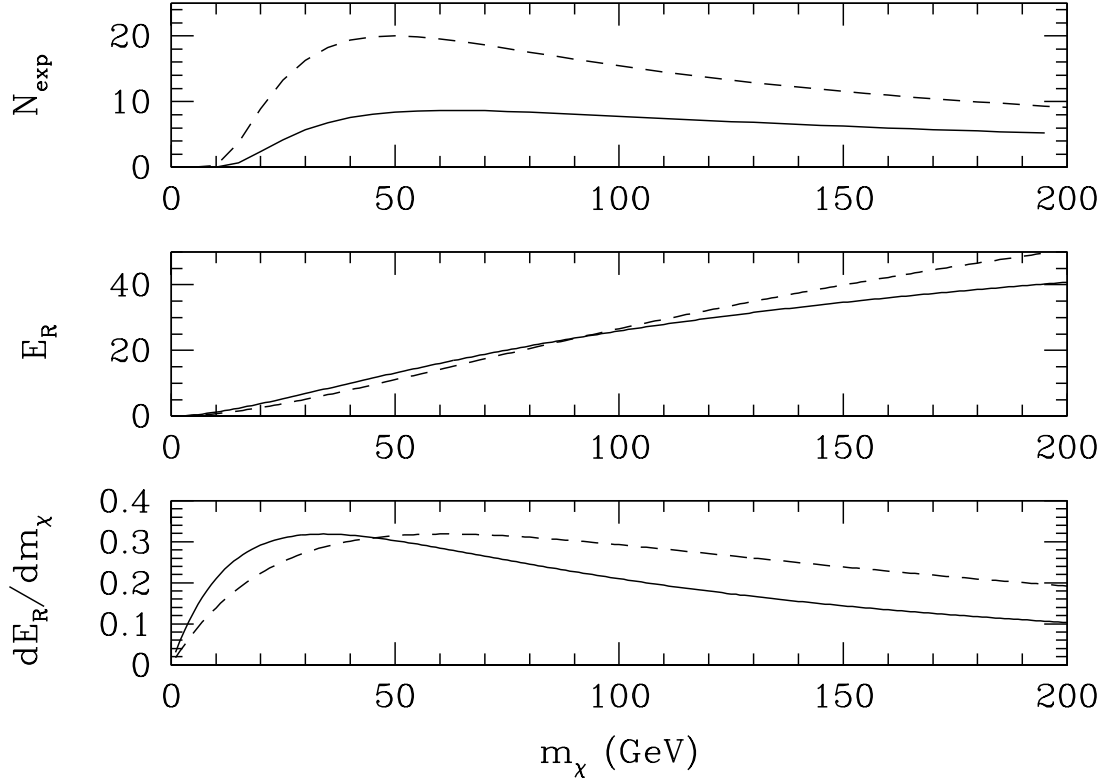} 
\end{center}  
\begin{center} 
  \caption{From top panel to bottom: the expected number of events,
    $N_{\rm exp}$, for a ${\cal E}= 3 \times 10^{3} \, {\rm kg \, day}$
    exposure (with $\sigma_{\rm p}=10^{-8} \, {\rm pb}$), the
    characteristic energy scale, $E_{\rm R}$, and the variation of the
    characteristic energy scale with mass, ${\rm d E_{\rm R}}/{\rm d}
    m_{\chi}$, as a function of WIMP mass for a Ge detector (solid line) and
    a Xe detector (dashed). For $N_{\rm exp}$ the 
    the energy thresholds of the current CDMS II~\cite{CDMS} and 
Xenon10~\cite{Xenon10} experiments
 have been used: $E_{\rm th} = 10 \, {\rm keV}$ and  $4.5 \,
    {\rm keV}$ for Ge and Xe respectively.}
\label{ernexp}  
\end{center} 
\end{figure}

The characteristic energy, $E_{\rm R}$, depends on the WIMP mass,
$m_{\chi}$, and the mass of the target nuclei, $m_{\rm A}$. For WIMPs
which are light compared with the target nuclei, $m_{\chi} \ll m_{\rm
  A}$, $E_{\rm R} \propto m_{\chi}^2/m_{\rm A}$, while for heavy
WIMPs, $m_{\chi} \gg m_{\rm A}$, $E_{\rm R} \sim {\rm const}$. In
words, for light WIMPs the energy spectrum is strongly dependent on
the WIMP mass while for heavy WIMPs the dependence on the WIMP mass is
far weaker. Consequently it will be easier to measure the mass of
light (compared with the target nuclei) than heavy WIMPs.  Since the
experiments which currently have the greatest sensitivity are composed
of Ge and Xe (CDMS II~\cite{CDMS} and Xenon10~\cite{Xenon10}
respectively) we focus on these targets.  Fig.~\ref{ernexp} shows the
dependence of the characteristic energy, $E_{\rm R}$, and ${\rm
  d}E_{\rm R}/ {\rm d} m_{\chi}$ on the WIMP mass for Ge and Xe. For
$m_{\chi} < (>) \sim 50 \, {\rm GeV}$ $E_{\rm R}$ varies more strongly
with $m_{\chi}$ for Ge (Xe), reflecting the asymptotic WIMP mass
dependences of the expression for $E_{\rm R}$. This suggests that
light (heavy) target nuclei will be better suited to determining the
mass of light (heavy) WIMPs (see however Sec.~\ref{Xesec}). 
The detector energy threshold 
will also come into play, in particular for small exposures, as the
expected number of events depends on the energy threshold. The
expected number of events, $N_{\rm exp}$, for an exposure of
 ${\cal E} = 3 \times
10^{3} \, {\rm kg \, day}$ is also shown in fig.~\ref{ernexp} as a function
of WIMP mass for Ge and Xe detectors with $E_{\rm th}= 10 \, {\rm
  keV}$~\cite{CDMS,SuperCDMS} and $4.5 \, {\rm keV}$~\cite{Xenon10}
respectively. With these thresholds $N_{\rm exp}$ is larger (by a
factor of order $\sim 2$ for $m_{\chi} \approx 50 \, {\rm GeV}$) for
Xe than for Ge.  This indicates that the relative capabilities of
detectors to determine the WIMP mass will depend not only on the WIMP
and target masses, but also on their energy thresholds.

\subsection{Monte Carlo simulations}

We use Monte Carlo simulations to examine, for a range of detector
configurations and input WIMP masses, how well the WIMP mass could be
determined from the energies of observed WIMP nuclear recoil events.

We estimate the WIMP mass and cross-section by maximizing the extended
likelihood function (which takes into account the fact that the number of
events observed in a given experiment is not fixed),
e.g. Ref.~\cite{cowan}:
\begin{equation} 
  L= \frac{\lambda^{N_{\rm expt}} \exp{(-\lambda)}}{N_{\rm expt}!}
    \Pi_{i=1}^{N_{\rm expt}} f(E_{\rm i}) \,.
\end{equation}
Here $N_{\rm expt}$ is the number of events observed, $E_{\rm i} \,
(i=1,..., N_{\rm expt})$ are the energies of the events observed,
$f(E)$ is the normalized differential event rate and $\lambda =
{\cal E} \int_{E_{th}}^{\infty} ({\rm d} R/{\rm d} E) \, {\rm d} E$ is
the mean number of events where ${\cal E}$ is the detector exposure
(which has dimensions of mass times time) and $E_{\rm th}$ is the
threshold energy.  We calculate the probability distribution of the
maximum likelihood estimators of the WIMP mass and cross-section, for
each detector configuration and input WIMP mass, by simulating
$10^{4}$ experiments. We first calculate the expected number of
events, $\lambda_{\rm in}$, from the input energy spectrum. The actual
number of events for a given experiment, $N_{\rm expt}$, is drawn from
a Poisson distribution with mean $\lambda_{\rm in}$. We Monte Carlo
generate $N_{\rm expt}$ events from the input energy spectrum, from
which the maximum likelihood mass and cross-section for that
experiment are calculated.  Finally we find the (two-sided) $68\%$ and
$95 \%$ confidence limits on the WIMP mass from the maximum likelihood masses.

\section{Results}
\label{results}

In Sec.~\ref{Geres} we investigate the mass limits for a a SuperCDMS
like Ge detector~\cite{SuperCDMS} and their dependence on the detector
energy threshold, maximum energy and exposure, and the WIMP
cross-section.  In Sec.~\ref{Xesec} we compare the Ge mass limits with
those for a Xe detector, before examining the effects of uncertainties in
the local circular speed and non-negligible background in
Secs.~\ref{vcsec} and~\ref{backsec} respectively.

\subsection{Germanium}
\label{Geres}

We begin, as in paper I~\cite{pap1}, by looking at a SuperCDMS like
detector~\cite{SuperCDMS}, composed of Ge with a nuclear recoil energy
threshold of $E_{\rm th} = 10 \, {\rm keV}$. We assume that the
background event rate is negligible, as is expected for this
experiment located at SNOLab~\cite{SuperCDMS}, and that the energy
resolution is perfect~\footnote{Gaussian energy resolution, with full
  width at half maximum of order $1 \, {\rm keV}$~\cite{SuperCDMS},
  does not affect the WIMP parameters extracted from the energy
  spectrum~\cite{pap1}.}.  For simplicity we assume that the nuclear recoil
detection efficiency is independent of energy. The energy dependence
of the efficiency of the current CDMS II experiment is relatively small
(it increases from $\sim 0.22$ at $E=E_{\rm th}=10 \, {\rm keV}$
to $\sim 0.30$ at $E=15 \, {\rm keV}$ and then remains roughly constant).
For further discussion of these assumptions see Ref.~\cite{pap1}. 

\begin{figure}  
\begin{center}  
\epsfxsize=6.in  
\epsfbox{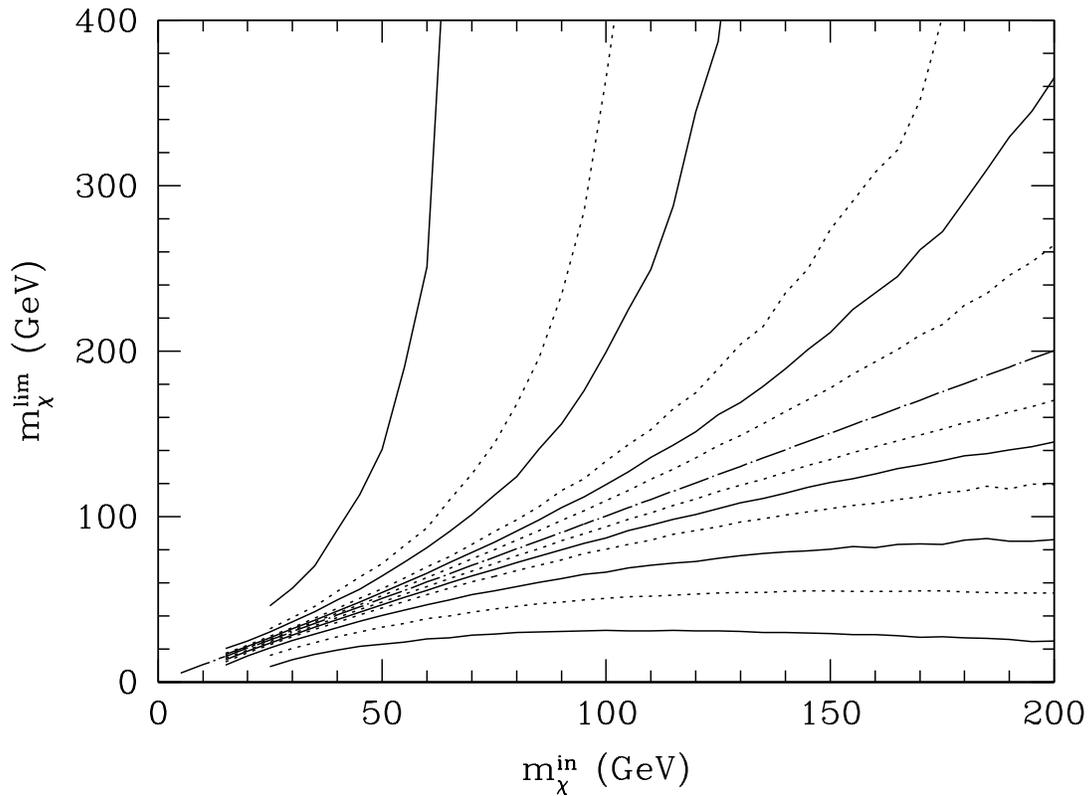} 
\end{center}  
\begin{center} 
  \caption{Limits on the WIMP mass, $m_{\chi}^{\rm lim}$,
 as a function of the input WIMP mass, $m_{\chi}^{\rm in}$,
 for the benchmark Ge detector 
($E_{\rm th} = 10 \, {\rm keV}$, perfect energy resolution,
    no upper limit on energy of events detected and zero background) 
for exposures ${\cal
      E}= 3 \times 10^{3}, 3 \times 10^{4}$ and $3 \times 10^{5} \,
    {\rm kg \, day}$ and input cross-section $\sigma_{\rm p} = 10^{-8}
    \, {\rm pb}$. The dot-dashed line is the input mass and the solid
    (dotted) lines are the 95\% (68\%) confidence limits.}
\label{Gevarym}  
\end{center} 
\end{figure}

\begin{figure}  
\begin{center}  
\epsfxsize=6.in  
\epsfbox{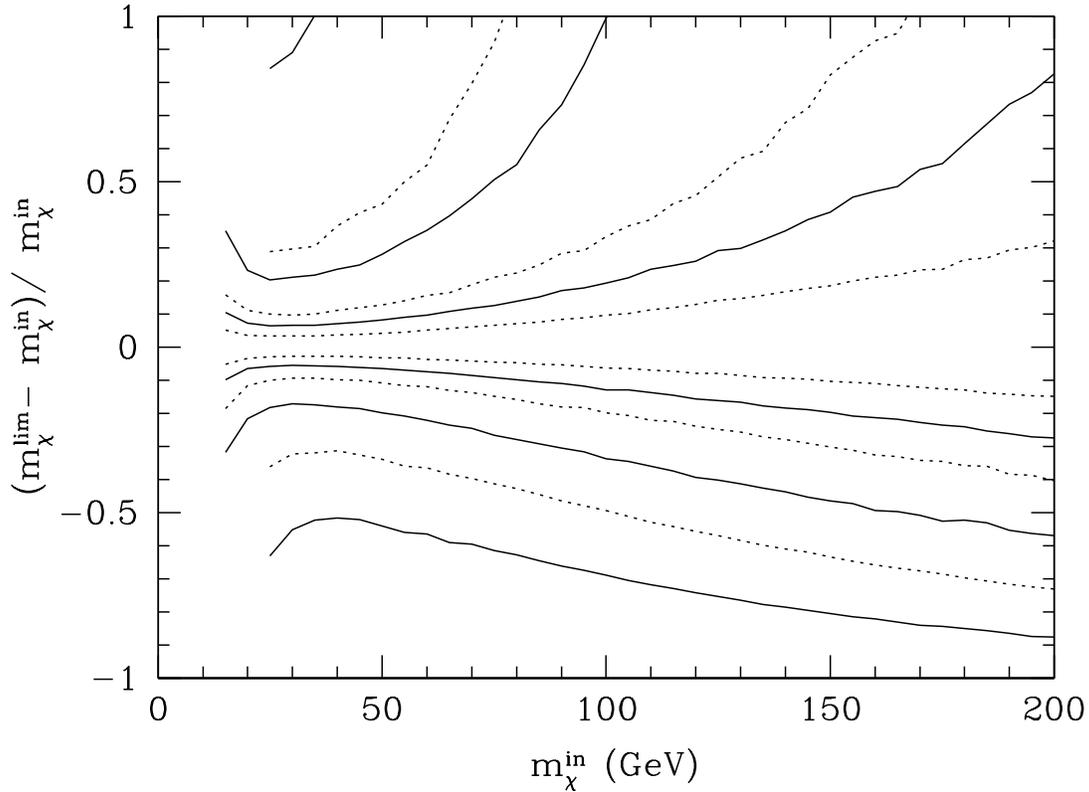} 
\end{center}  
\begin{center} 
  \caption{As fig.~\ref{Gevarym}, but with the fractional deviation
of the WIMP mass limits from the input mass, $(m_{\chi}^{\rm lim}-m_{\chi}^{\rm
      in})/m_{\chi}^{\rm in}$, plotted.}
\label{Gefrac}  
\end{center} 
\end{figure}

We consider fiducial efficiency weighted exposures~\footnote{For
  brevity we subsequently refer to this as simply the exposure.}
${\cal E}=3 \times 10^{3}$, $3 \times 10^{4}$ and $3 \times 10^{5} \,
{\rm kg \, day}$ which correspond, roughly, to a detector with mass
equal to that of the 3 proposed phases of SuperCDMS taking data for a
year with a $\sim 50\%$ detection efficiency~\footnote{The ~$50\%$
detection efficiency was chosen based on Ref~\cite{CDMSold}. The more recent
CDMS II analysis~\cite{CDMS} has a lower nuclear recoil acceptance,
$\sim 30\%$.}.

We use fiducial values for the detector energy threshold and
WIMP-proton cross-section of $E_{\rm th}= 10 \, {\rm keV}$ and
$\sigma_{\rm p} = 10^{-8} \, {\rm pb}$ but later,
in Secs.~\ref{ethsec} and \ref{sigmasec} respectively,
consider a range of values for these parameters. Note that this
fiducial cross-section is, given the recent limits from the
CDMS II~\cite{CDMS} and Xenon10~\cite{Xenon10} experiments, an order of
magnitude smaller than that used in Paper I~\cite{pap1}. We use the
standard values for the local circular speed and WIMP density, $v_{\rm
  c} = 220 \, {\rm km \, s }^{-1}$ and $\rho_{\chi} = 0.3 \, {\rm GeV
  \, cm}^{-3}$ respectively. We examine the effect of uncertainties
in the local circular speed in Sec.~\ref{vcsec}. The local WIMP density
only affects the amplitude, and not the shape, of the energy spectrum.
Therefore it only affects the WIMP mass determination indirectly, through
the number of events detected.
We note that the limits from our idealized
simulated detector are likely to be better than those achievable in
reality by a real detector.

In fig.~\ref{Gevarym} we plot the 68$\%$ and $95\%$ confidence limits
on the WIMP mass, $m_{\chi}^{\rm lim}$, for the fiducial detector
configuration as a function of the input WIMP mass, $m_{\chi}^{\rm
  in}$. Here, and throughout, the limits terminate when there is a $>
5\%$ probability that an experiment will detect no events.
Fig.~\ref{Gefrac} uses the same data, but shows the fractional limits
on the WIMP mass, $(m_{\chi}^{\rm lim}-m_{\chi}^{\rm
  in})/m_{\chi}^{\rm in}$. With exposures of ${\cal E}= 3
\times 10^{4}$ and $3 \times 10^{5} \, {\rm kg \, day}$ it would be
possible, with this detector configuration, to measure the mass of a
light [$m_{\chi} \sim {\cal O}(50 \, {\rm GeV})$] WIMP with an accuracy
of roughly $25\%$ and $10\%$ respectively.  These numbers, and the
upper limits in particular, increase with increasing WIMP mass, and
for heavy WIMPs ($m_{\chi} \gg 100 \, {\rm GeV}$) even with a large
exposure it will only be possible to place a lower limit on the mass.
For very light WIMPs, $m_{\chi} < {\cal O}(20 \, {\rm GeV})$, the
number of events above the detector energy threshold would be too
small to allow the mass to be measured accurately.

\subsubsection{Energy threshold}
\label{ethsec}
 
We now examine the effects of varying the energy threshold, $E_{\rm
  th}$ (for the fiducial detector configuration and WIMP properties
described above).  Table 1 contains the $95\%$ confidence limits on
the fractional deviation of the WIMP mass from the input WIMP mass for
input WIMP masses of $m_{\chi}^{\rm in}= 50, \, 100$ and $200 \, {\rm
  GeV}$, energy thresholds $E_{\rm th}= 0, \, 10$ and $20 \, {\rm
  keV}$ and an exposure of ${\cal E} = 3 \times 10^{5} \, {\rm kg \,
  day}$.
  
As the energy threshold is increased the expected number of events
decreases. The smaller range of recoil energies also reduces the
accuracy with which the characteristic scale of the energy spectrum,
$E_{\rm R}$, and hence the WIMP mass can be determined. The effect of
varying $E_{\rm th}$ is smallest for intermediate WIMP masses; for
$m_{\chi}= 50 \, {\rm GeV}$ (and light WIMPs in general) the small
$E_{\rm R}$ means that expected number of events decreases rapidly as
the energy threshold is increased, while for $m_{\chi}= 200 \, {\rm
  GeV}$ (and heavy WIMPs in general) the large $E_{\rm R}$, and
flatter energy spectrum, means that the smaller range of recoil
energies reduces the accuracy with which $E_{\rm R}$ can be measured.

\begin{table}[t]
\begin{center}
  \caption{Dependence of the $95\%$ fractional confidence limits on
     the WIMP mass, 
 $(m_{\chi}^{\rm lim}- m_{\chi})/m_{\chi}$, on the
    energy threshold, $E_{\rm th}$, for the benchmark Ge detector, for
    input WIMP masses $m_{\chi}^{\rm in} = 50,\, 100$ and $200 \, {\rm
      GeV}$ and exposure ${\cal E}= 3 \times 10^{5} \, {\rm kg \,
      day}$.}
\label{ethtable}
\medskip
\begin{tabular}{||c||c||c||c||}
\hline
 $E_{\rm th}$ \, (keV) & \multicolumn{3}{|c|}{$m_{\chi}^{\rm in}$ \, (GeV)} \\
  &    50 & 100 & 200 \\
\hline
\hline
0 & -0.064 \,, +0.043  & -0.12 \,, +0.14 & -0.25 \,, +0.63  \\ 
\hline
10 & -0.082 \,, +0.057  & -0.13 \,, +0.18 & -0.27 \,, +0.81 \\ 
\hline
20 & -0.11\,, +0.073 & -0.16 \,, +0.22 & -0.31 \,, +1.1 \\ 
\hline
\hline
\end{tabular}
\end{center}
\end{table}

\subsubsection{Maximum energy}

\begin{figure}  
\begin{center}  
\epsfxsize=6.in  
\epsfbox{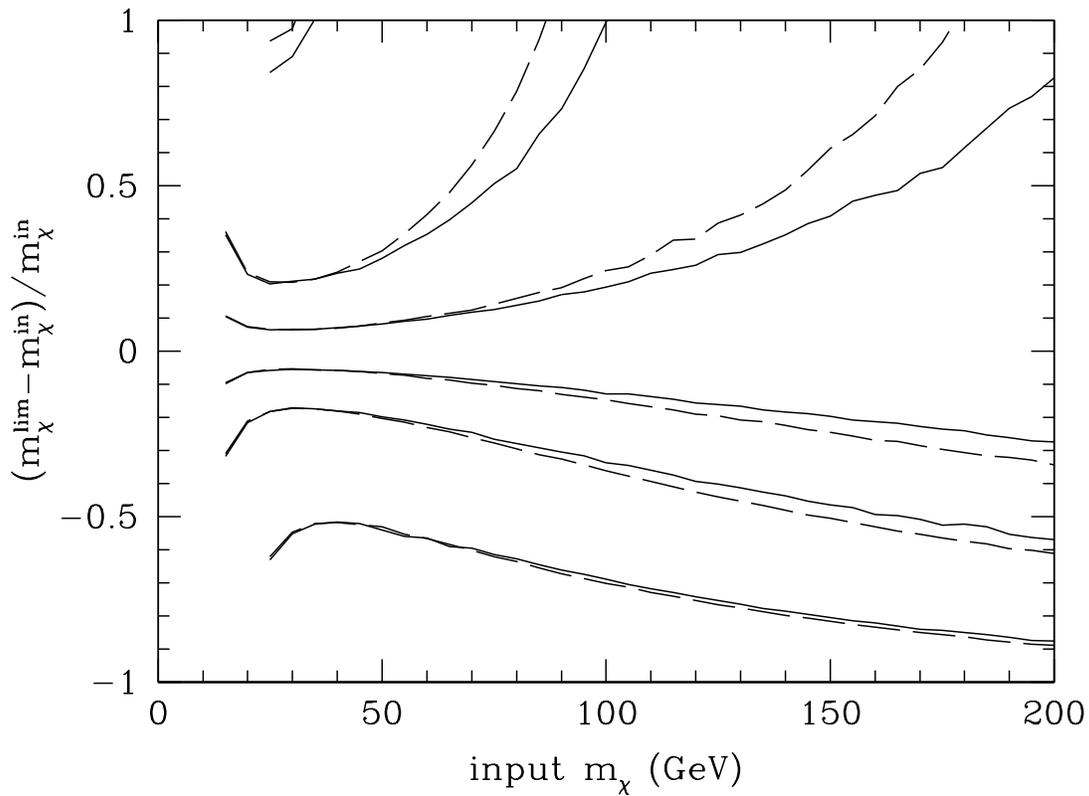} 
\end{center}  
\begin{center} 
  \caption{Fractional mass limits as a function of input mass for the
    fiducial detector configuration, with no limit on the energy of
    recoils which can be detected, (solid lines as before) and for a
    maximum energy $E_{\rm max}=100 \, {\rm GeV}$ (long dashed). For
    clarity only the $95\%$ confidence limits are displayed in this
    figure.}
\label{Geemax100}  
\end{center} 
\end{figure}

We have previously assumed that recoil events of all energies above
the threshold energy can be detected.  In real experiments there will
be a maximum energy, $E_{\rm max}$, above which recoils are not
detected/analysed. For instance for CDMS II~\cite{CDMS} $E_{\rm
  max}=100 \, {\rm keV}$. Fig.~\ref{Geemax100} compares the fractional
mass limits for $E_{\rm max}=100 \, {\rm keV}$ with those found
previously assuming no upper limit. The difference is very small for
light WIMPs [$m_{\chi} < {\cal O}(50 \, {\rm GeV})$] increasing with
increasing $m_{\chi}$ to ${\cal O}(10 \%)$ for $m_{\chi} \sim {\cal
  O}(200 \, {\rm GeV})$ and ${\cal E}= 3 \times 10^{5} \, {\rm kg \,
  day}$. This reflects the fact that for light WIMPs the differential
event rate above $E_{\rm max}$ is essentially negligible, however this
is not the case for heavier WIMPs and finite $E_{\rm max}$ reduces the
accuracy with which the characteristic energy scale of the spectrum
can be measured.

\subsubsection{Exposure}

\begin{figure}  
\begin{center}  
\epsfxsize=6.in  
\epsfbox{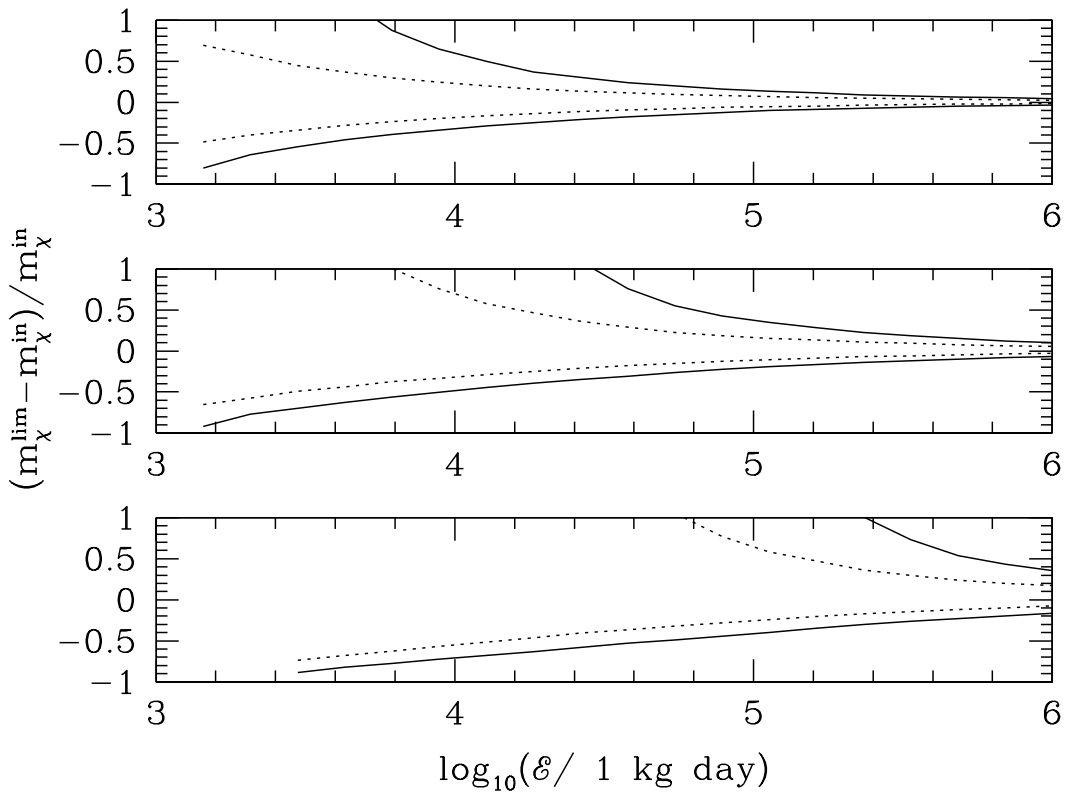} 
\end{center}  
\begin{center} 
  \caption{Fractional mass limits as a function of exposure, ${\cal E}$, for the
    benchmark Ge detector and $\sigma_{\rm p}=10^{-8} \, {\rm pb}$,
for, from top to bottom, $m_{\chi}^{\rm in}=50, \,
    100$ and $200\,  {\rm GeV}$.}
\label{Geexpos}  
\end{center} 
\end{figure}

Fig.~\ref{Geexpos} shows the fractional limits on the WIMP mass as a
function of the exposure, ${\cal E}$, for input WIMP masses of
$m_{\chi}^{\rm in}= 50, \, 100$ and $200 \, {\rm GeV}$, for the
fiducial detector and $\sigma_{\rm p}=10^{-8} \, {\rm pb}$. As the
exposure is increased the mass limits (and in particular the $95\%$
upper confidence limit) improve, initially rapidly and then more slowly
(reflecting the fact that the expected number of events is directly
proportional to the exposure). An accuracy of $\sim \pm 10\%$ in the
determination of the WIMP mass can be achieved with an exposure ${\cal
  E}= 10^{5} \, (10^{6}) \, {\rm kg \, day} $ for $m_{\chi}^{\rm in} =
50 \, (100) \, {\rm GeV}$.  For $m_{\chi}^{\rm in} = 200 \, {\rm GeV}$
even ${\cal E}= 10^{6} \, {\rm kg \, day}$ would not be sufficient to
achieve ${\cal O}(10\%)$ precision.

\subsubsection{Cross-section}
\label{sigmasec}

The expected number of events is directly proportional to both the
cross-section and exposure. Varying the cross-section is therefore
very similar to varying the exposure (and hence we do not display a plot
of the limits for varying $\sigma_{\rm p}$).

Unsurprisingly the accuracy with which it will be possible to
determine the WIMP mass depends sensitively on the underlying
cross-section. For instance if $\sigma_{\rm p}= 10^{-9} \, {\rm pb} $,
with an exposure of ${\cal E} = 3 \times 10^{5} \, {\rm kg \, day}$ it
will be possible to measure the mass with an accuracy of $\sim \pm 20
\% \, ({}_{-30\%}^{ + 100\%})$ for $m_{\chi}^{\rm in} = 50 \, (100) \,
{\rm GeV}$. If $\sigma_{\rm p}= 10^{-10} \, {\rm pb} $, even for
$m_{\chi}^{\rm in} = 50 \, {\rm GeV}$ with an exposure of ${\cal E} =
3 \times 10^{5} \, {\rm kg \, day}$ it will only be possible to
determine the WIMP mass to within a factor of a few and for more
massive WIMPs it will only be possible to place a lower limit on the
mass.

\subsection{Xenon}
\label{Xesec}
\begin{figure}  
\begin{center}  
\epsfxsize=6.in  
\epsfbox{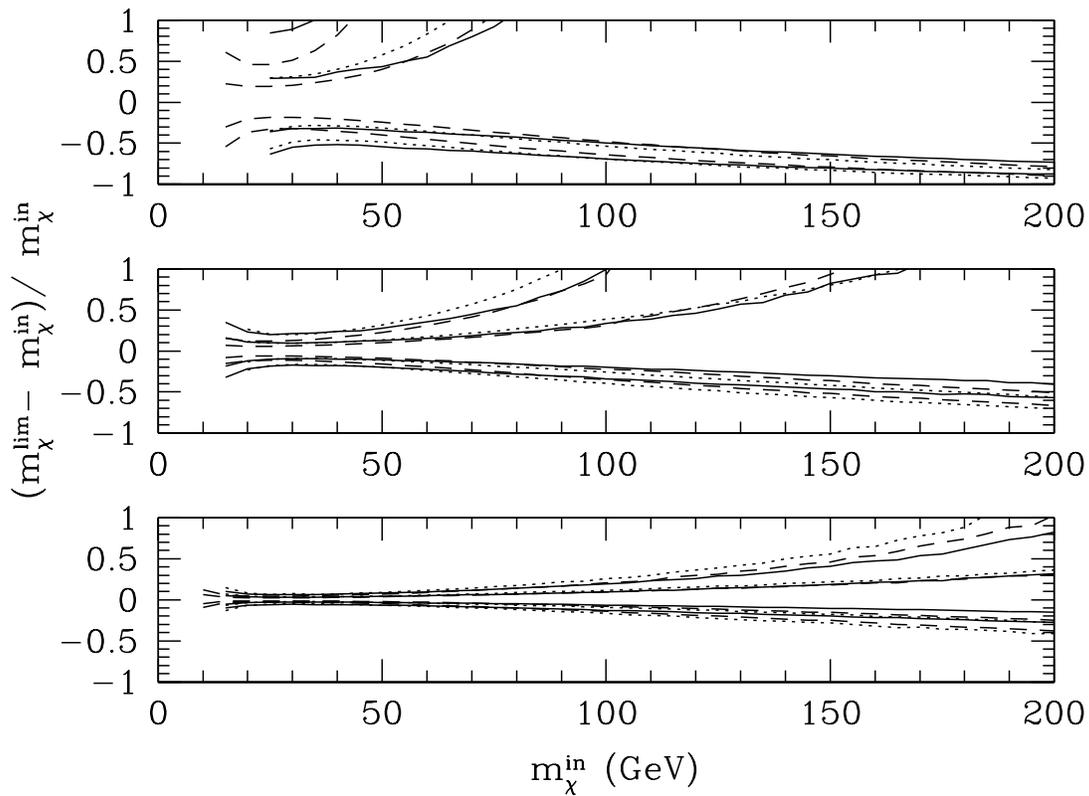} 
\end{center}  
\begin{center} 
  \caption{Fractional mass limits  for
    the fiducial Ge detector with $E_{\rm th}= 10 \, {\rm keV}$
 (solid lines for both $68\%$ and $95\%$ confidence limits) and for a 
Xe detector with
    $E_{\rm th}=4.5 \, $ and $10 \, {\rm keV}$ (dashed and dotted
    lines respectively) for (from top to bottom) ${\cal E} = 3 \times
    10^{3}, 3 \times 10^{4}$ and $3 \times 10^{5}\, {\rm kg \, day}$.
}
\label{Xefracpanel}  
\end{center} 
\end{figure}

\begin{figure}  
\begin{center}  
\epsfxsize=6.in  
\epsfbox{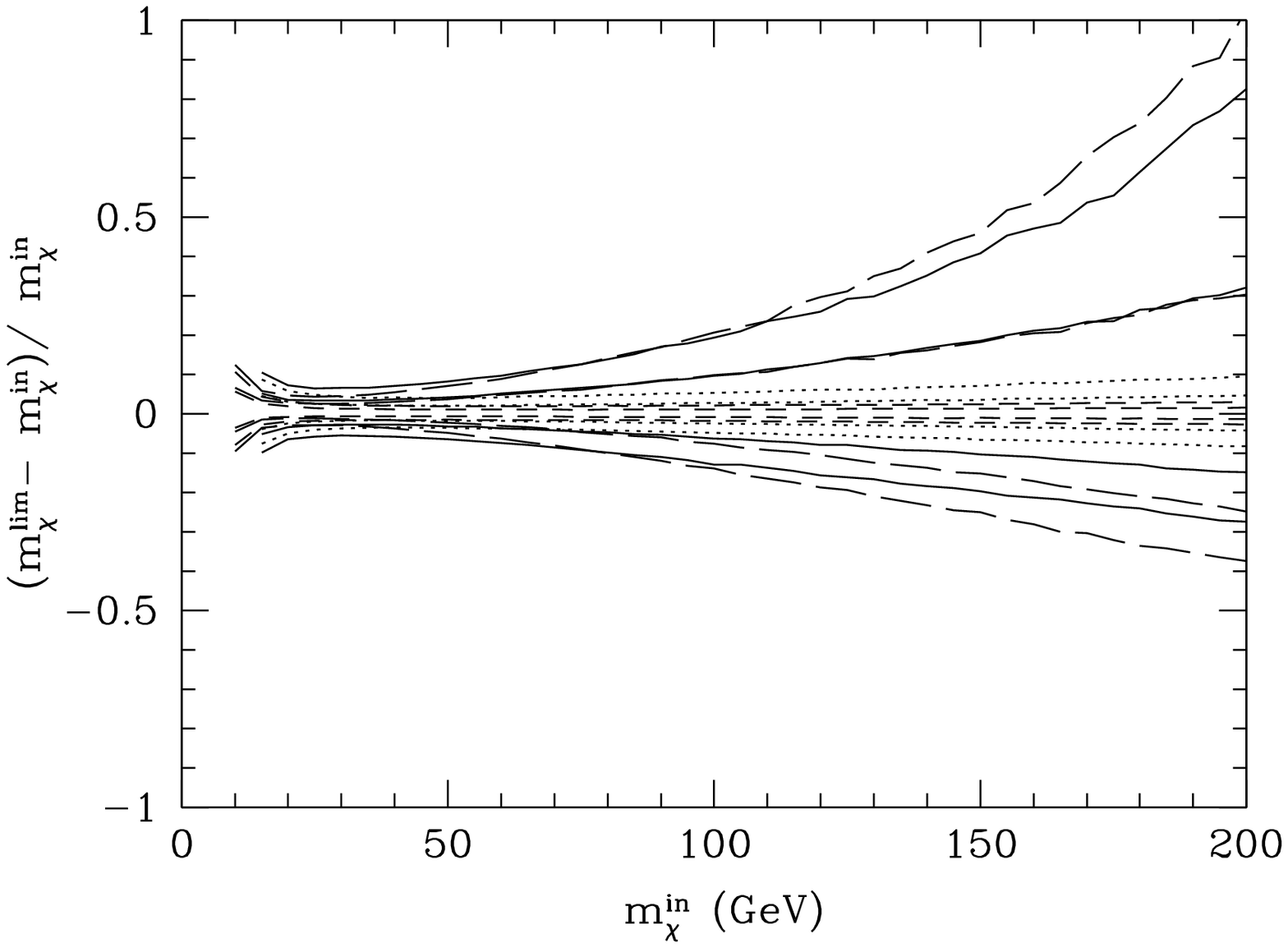} 
\end{center}  
\begin{center} 
  \caption{ A comparison of the fractional mass limits obtained, for
    Ge and Xe with an exposure ${\cal E}= 3 \times 10^{5} \, {\rm kg
      \, day}$, with and without the form factor. The solid and long
    dashed lines are for the fiducial Ge detector with $E_{\rm th}= 10
    \, {\rm keV}$ and for a Xe detector with $E_{\rm th}=4.5 \, {\rm
      keV}$ respectively, as in the bottom panel of
    Fig.~\ref{Xefracpanel}. The dotted and short dashed lines are for
    the same Ge and Xe detectors, with the form factor artificially 
set to unity,
    $F(E)=1$.}
\label{GeXeFFb}  
\end{center} 
\end{figure}

We now examine the dependence of the mass limits on the detector
target material.  In fig.~\ref{Xefracpanel} we compare the fractional
limits for the fiducial (Super-CDMS~\cite{SuperCDMS} like) detector
with $E_{\rm th}=10 \, {\rm keV}$ with those from a Xe detector with
$E_{\rm th}=4.5 \, {\rm keV}$ (c.f. Xenon10~\cite{Xenon10}) and $10 \,
{\rm keV}$. The difference in the mass limits for the different
detector configurations is largest for small exposures, where the
number of events observed is small. For light WIMPs, as found in
Sec.~\ref{ethsec}, the energy threshold is important, with the Xe
detector with $E_{\rm th}= 4.5 \, {\rm keV}$ doing significantly
better than both the fiducial Ge detector and also the Xe detector
with $E_{\rm th}= 10 \, {\rm keV}$. For heavier WIMPs, $m_{\chi} > 100
\, {\rm GeV}$, in contrast to the naive expectation from the
$m_{\chi}$ dependence of $E_{\rm R}$ (see Sec.~\ref{secdrde}), the Ge detector
produces slightly better limits than the Xe detector with $E_{\rm th}=
4.5 \, {\rm keV}$.  This is because of the rapid decrease of the Xe
form factor with increasing energy/momentum transfer. In
fig.~\ref{GeXeFFb} we compare, for a Ge detector with $E_{\rm th}= 10
\, {\rm keV}$ and a Xe detector with $E_{\rm th}= 4.5 \, {\rm keV}$,
the mass limits we found before with those which would be obtained if
the form factor were equal to unity for all energies,
$F(E)=1$. Without the form factor the mass limits are substantially
tighter for both Ge and Xe. This is because the event rate, in
particular at large energies, is increased.  The improvement in the
accuracy of the determination of the characteristic energy, $E_{\rm
  R}$, and hence $m_{\chi}$, is greater than if the exposure were
simply increased so as to increase the expected number of events.  For
instance for Xe, $m_{\chi}=200 \, {\rm GeV}$ and ${\cal E}= 3 \times
10^{5} \, {\rm kg \, day}$ without the form factor the one-$\sigma$
error on the WIMP mass is $\sim 3 \, {\rm GeV}$, while with the form
factor included if the exposure is increased so as to give the same
expected number of events the one-$\sigma$ error is $\sim 20 \, {\rm
  GeV}$.  This is because the greater relative abundance of large
energy recoils allows $E_{\rm R}$ to be determined more accurately.
With the form factor set to unity the mass limits for large WIMP
masses are significantly better for Xe than for Ge, as naively
expected from the WIMP mass dependence of $E_{\rm R}$.  However for
any real detector the rapid decrease of the Xe form factor with
increasing energy/momentum transfer means that, contrary to naive
expectations, the mass of heavy WIMPs can not be measured more
accurately with Xe than with Ge (assuming similar threshold
energies). This conclusion, see also recent discussion by Drees and
Shan~\cite{sd}, should also hold for any other heavy target.

\subsection{Varying input circular speed, $v_{\rm c}$}
\label{vcsec}

Up until now we have assumed that the local circular speed, $v_{\rm
  c}$, is known and equal to its standard value of $220 \, {\rm km
  \, s}^{-1}$.  There is in fact an uncertainty in $v_{\rm c}$ of
order $\pm 20 \, {\rm km \, s}^{-1}$~\cite{klb} and since $E_{\rm R}$
depends on both $m_{\chi}$ and $v_{\rm c}$ there is a degeneracy between
$m_{\chi}$ and $v_{\rm c}$~\cite{pap1}. Physically, the kinetic
energies of the incoming WIMPs depend on their mass and
velocities. For larger (smaller) $v_{\rm c}$ the incoming WIMPs have
larger (smaller) mean kinetic energies than assumed, resulting in
larger (smaller) maximum likelihood mass values.  This can be made more
quantitative by differentiating the expression for the characteristic
energy $E_{\rm R}$,~eq.~(\ref{er}):
\begin{equation}
    \frac{{\Delta} m_{\chi}}{m_{\chi}} = - [1+ (m_{\chi}/m_{\rm A})]
     \frac{\Delta v_{\rm c}}{v_{\rm c}}  \,.
\end{equation}
For an input WIMP mass of $m_{\chi}^{\rm in}=100\, {\rm GeV}$ and a $20 \, {\rm
  km \, s}^{-1}$ uncertainty in $v_{\rm c}$, this gives a $ \sim 20 \,
{\rm GeV}$ shift in the value of the WIMP mass determined.

\begin{figure}  
\begin{center}  
\epsfxsize=6.in  
\epsfbox{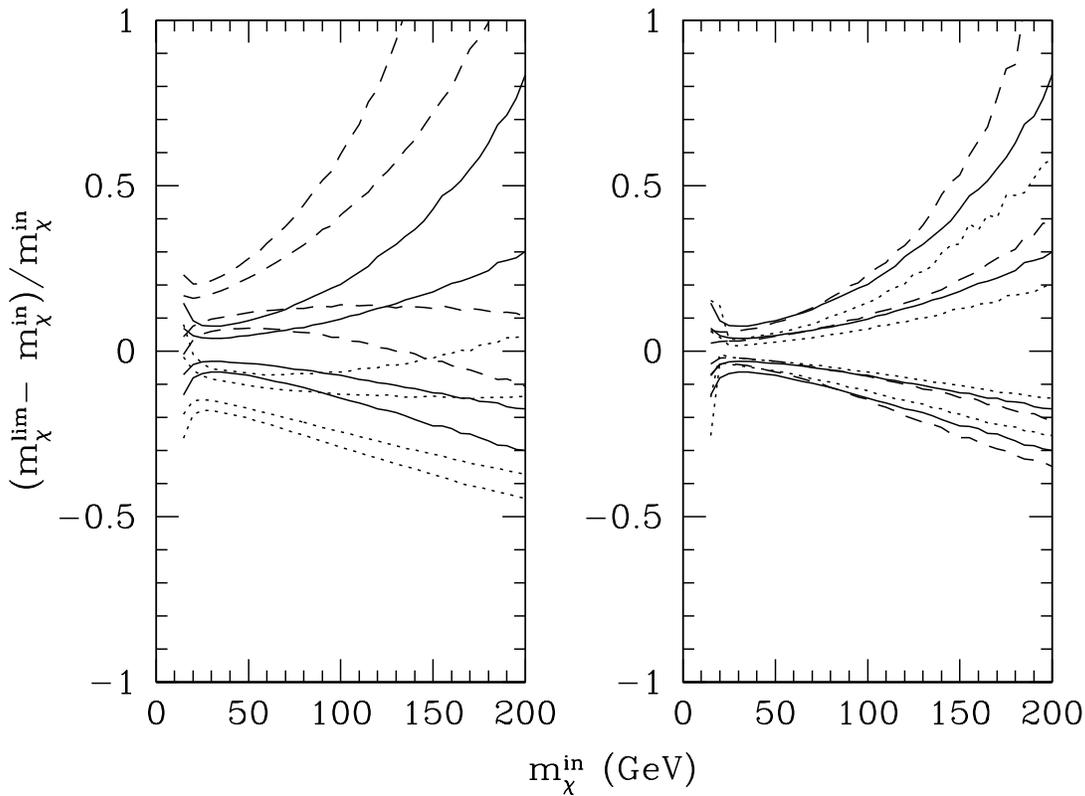} 
\end{center}  
\begin{center} 
  \caption{Fractional mass limits for the fiducial Ge detector
and ${\cal E}= 3 \times 10^{5} \, {\rm kg \, day}$ for varying circular
speed, $v_{\rm c}$. In the left panel $v_{\rm c}$ is fixed as $220 \, {\rm km \,
s}^{-1}$ in the likelihood analysis while in the right panel $v_{\rm c}$
is allowed to vary in the likelihood analysis (i.e. it is an additional fitting
parameter). Solid, dotted and dashed lines are for $v_{\rm c}^{\rm in} = 220, \,
200$ and $240 \, {\rm km \, s}^{-1}$ respectively.
}
\label{vcfig}  
\end{center} 
\end{figure}

In fig.~\ref{vcfig} we plot the fractional mass limits for input
circular speeds $v_{\rm c}^{\rm in} = 200,\, 220$ and $240 \, {\rm km
  \ s}^{-1}$ for ${\cal E} = 3 \times 10^{5} \, {\rm kg \, day}$. We
carry out the likelihood analysis twice, once with $v_{\rm c}$ fixed
at $220 \, {\rm km \, s}^{-1}$ and once with $v_{\rm c}$ as an additional
variable parameter. As discussed in Sec.~\ref{secdrde} in this section
we use the fitting function, eq.~(\ref{drde0}), for both the input energy
spectrum and the likelihood analysis as it is not computationally
feasible to carry out the full calculation of the energy spectrum for
each value of $v_{\rm c}$ considered during the likelihood
analysis.

When the underlying value of $v_{\rm c}^{\rm in}$ is different from
the (fixed) value used in the likelihood analysis there is, as
expected, a significant systematic error in the determination of the
WIMP mass.  For deviations of $\pm 20 \, {\rm km \, s}^{-1}$ in the
underlying value of $v_{\rm c}$ this systematic error increases with
increasing $m_{\chi}^{\rm in}$ from $\sim 10\% $ for small
$m_{\chi}^{\rm in}$ to $\sim 40\%$ for $m_{\chi} \approx 200 \, {\rm
  GeV}$. The limits are however asymmetric, with the systematic error
in the upper limits being substantially larger for $v_{\rm c}^{\rm
  in}=200 \, {\rm km \, s}^{-1}$. Allowing the value of $v_{\rm c}$ to
vary in the likelihood analysis substantially reduces the error, but
there still appears to be a small systematic shift in the mass limits.

\subsection{Backgrounds}
\label{backsec}

\begin{figure}  
\begin{center}  
\epsfxsize=6.in  
\epsfbox{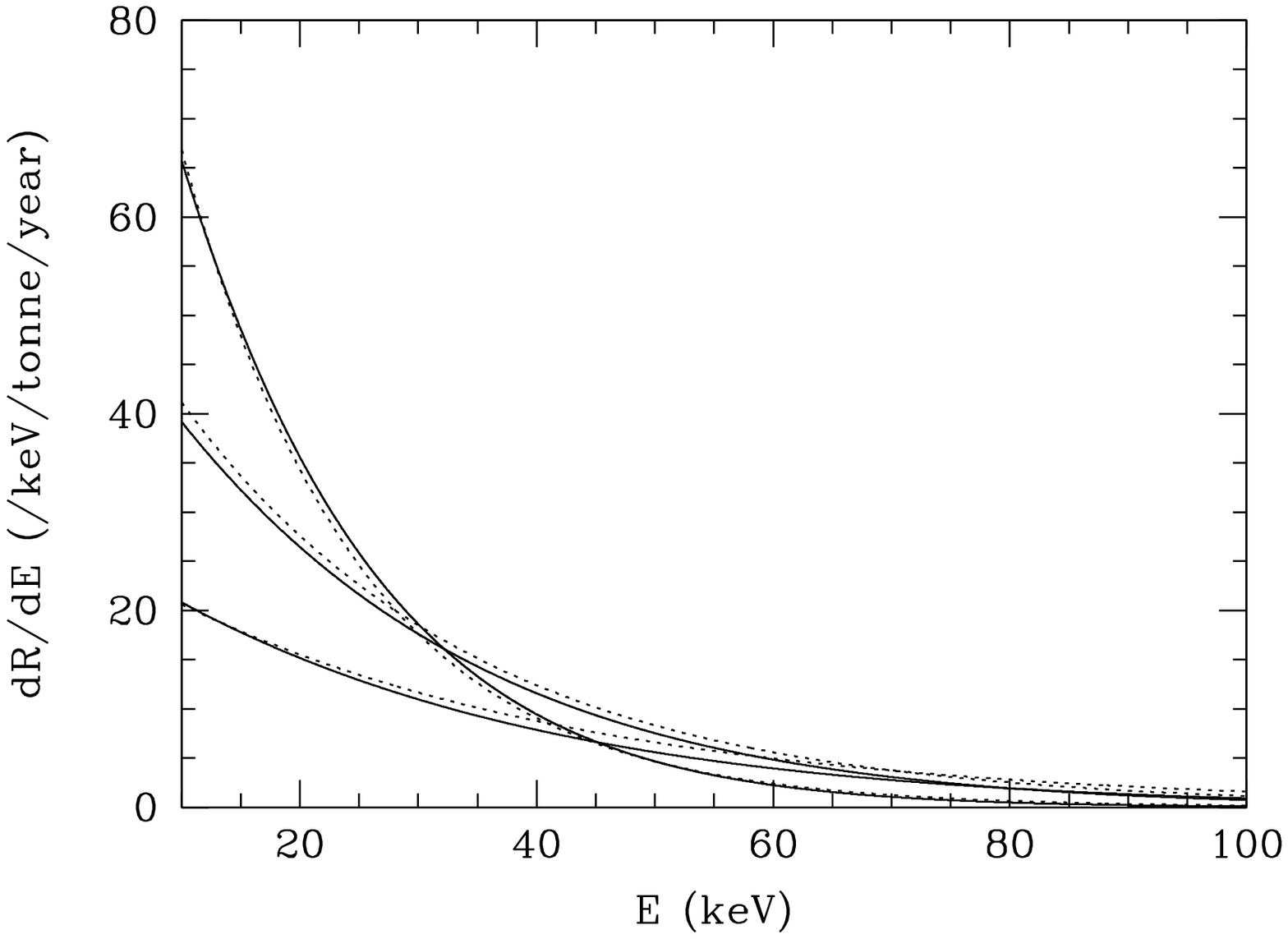} 
\end{center}  
\begin{center} 
  \caption{
Solid lines are the differential energy spectra including
the form factor, ${\rm d} R/ {\rm d} E$, for
    (from top to bottom at $E=0 \, {\rm keV}$) WIMPs with
 $m_{\chi} = 50, \, 100$ and $200 \, {\rm GeV}$ (solid lines). The dotted
lines are exponential background spectra with (from top to bottom)
$b_{\rm r}= 1000 \ {\rm tonne}^{-1} {\rm year}^{-1}$ \& $E_{\rm b}=15 \, 
{\rm keV}$,
$b_{\rm r}= 1000 \ {\rm tonne}^{-1} {\rm year}^{-1}$ \& $E_{\rm b}=25 \, {\rm keV}$
and $b_{\rm r}= 670 \ {\rm tonne}^{-1} {\rm year}^{-1}$ \& $E_{\rm b}=35 \, {\rm keV}$
  (dotted lines).
The parameters of the exponential background spectra have been chosen
to demonstrate that, even when the form factor is included, the WIMP
recoil spectra are close to exponential and could, in principle, be
mimicked by an exponential background.}
\label{drdemer}  
\end{center} 
\end{figure}

While future experiments aim to have negligible backgrounds
(e.g. Ref.~\cite{SuperCDMS}), non-negligible neutron backgrounds would
lead to errors in the determination of the WIMP mass. The size of the
errors will depend on the amplitude and shape of the background
spectrum. In particular if the background spectrum is exponential it
can closely mimic the shape of a WIMP recoil spectrum (see
fig.~\ref{drdemer}).

Motivated by simulations of the neutron background in various
dark matter detectors~\cite{nb}
we consider two forms for the background
\begin{enumerate}
\item{A flat background energy spectrum from $E_{\rm th}=10 \, {\rm keV}$
to $E_{\rm max}=100 \, {\rm keV}$~\footnote{In this section we assume
that only recoils up to $E_{\rm max}$ are detected.}
, parametrised by the total background rate,
$b_{\rm r}$, per tonne year.}
\item{A exponential energy spectrum from $E_{\rm th}=10 \, {\rm keV}$
to $E_{\rm max}=100 \, {\rm keV}$, parametrised by the total background rate,
$b_{\rm r}$, per tonne year, and the characteristic background energy scale, $E_{\rm b}$:
\begin{equation}
\left(\frac{{\rm d} R}{{\rm d} E} \right)_{\rm back} =
\left(\frac{{\rm d} R}{{\rm d} E} \right)_{E=0} \exp{[-(E/E_{\rm b})]} \,,
\end{equation}
where
\begin{equation}
b_{\rm r}= \int_{E_{\rm th}}^{E_{\rm max}}
\left(\frac{{\rm d} R}{{\rm d} E} \right)_{\rm back} \,.
\end{equation}}
\end{enumerate}

\begin{figure}  
\begin{center}  
\epsfxsize=6.in  
\epsfbox{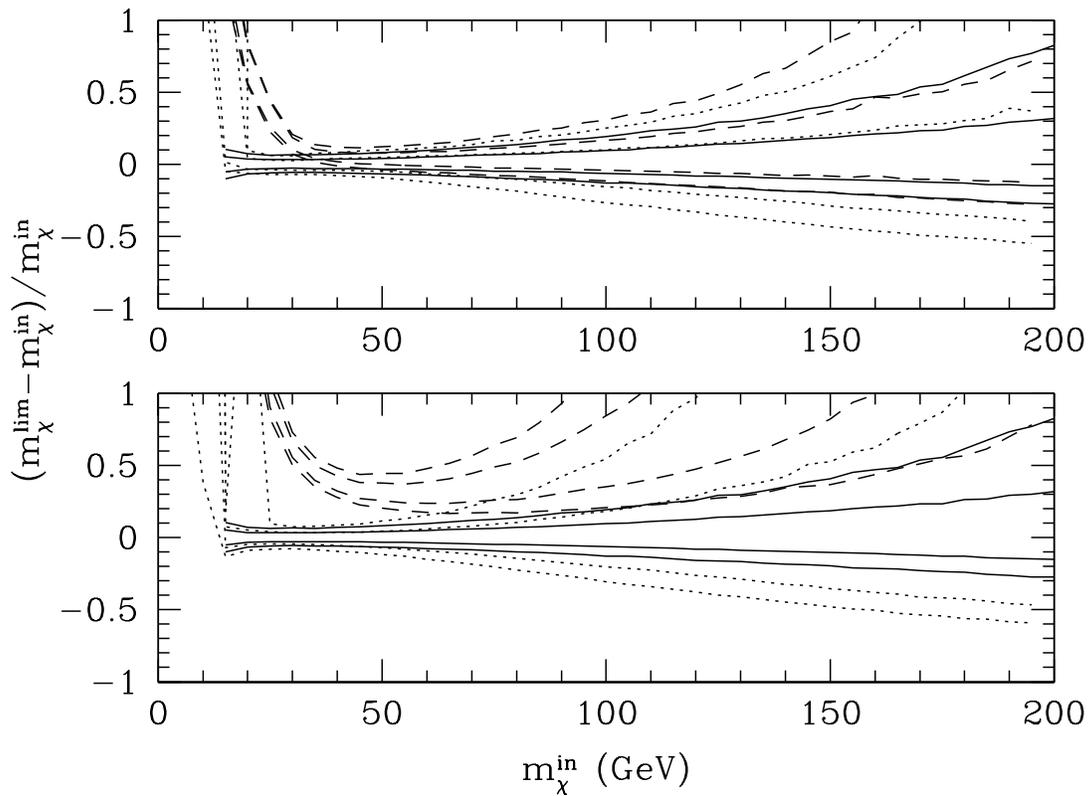} 
\end{center}  
\begin{center} 
  \caption{Fractional mass limits
in the presence of a background with a flat energy spectrum and total
rate $b_{\rm r}= 10$ and  $100 \, {\rm tonne}^{-1} \, {\rm year}^{-1}$ 
(top and bottom 
panels respectively). The dashed (dotted) lines are for when the background
event rate is not (is) included in the likelihood analysis. 
The solid lines are the confidence limits for zero background.
The fiducial Ge detector configuration with $E_{\rm max} = 100 \, {\rm keV}$,
${\cal E}= 3
\times 10^{5} \, {\rm kg \, day}$ and $\sigma_{\rm p}=10^{-8} \, {\rm
  pb}$ is used.}
\label{back1rat}  
\end{center} 
\end{figure}

The limits on the WIMP mass for an exposure of $ {\cal E}= 3 \times
10^{5} \, {\rm kg \, day}$ and $\sigma_{\rm p}=10^{-8} \, {\rm pb}$ in
the presence of a flat background with $b_{\rm r}= 10$ and $100 \,
{\rm tonne}^{-1} \, {\rm year}^{-1}$~\footnote{Note that the
  backgrounds in real experiments are not expected to be this large.}
are displayed in fig.~\ref{back1rat}. We carry out the likelihood
analysis of the WIMP parameters twice, firstly neglecting the
background and then including the background rate as an additional
parameter. Neglecting the background leads to a systematic
over-estimate of the WIMP mass, since the flat background increases
the event rate at large $E$ relative to that at small $E$, so that the
best fit energy spectrum has larger $E_{\rm R}$, or equivalently
larger $m_{\chi}$. Including the background rate in the likelihood
analysis avoids the systematic error but, inevitably, leads to larger
statistical error in the WIMP mass limits.  The fractional errors are
smallest for $m_{\chi} \sim {\cal O} (50 \, {\rm GeV})$ and increase
for smaller and larger WIMP masses. This is because for small WIMP
masses the background event rate is larger compared with the WIMP
event rate, while for large WIMP masses the WIMP energy spectrum
(exponential with large characteristic energy scale) is closer in
shape to the flat background spectrum.  The systematic (background not
included in analysis) and additional statistical (background rate
included) errors are both at least $10\%$ for $b_{\rm r}= 10 \, {\rm
  tonne}^{-1} \, {\rm year}^{-1}$. For $b_{\rm r}= 100 \, {\rm tonne}^{-1} \, {\rm
  year}^{-1}$ the minimum systematic error (for $m_{\chi} \sim 50 \,
{\rm GeV}$) is $\sim 30\%$, the shift in the lower limits when the
background is included in the likelihood analysis is not much larger
than for $b_{\rm r}= 10 \, {\rm tonne}^{-1} \, {\rm year}^{-1}$, however
the upper limits are increased significantly. 

\begin{figure}  
\begin{center}  
\epsfxsize=6.in  
\epsfbox{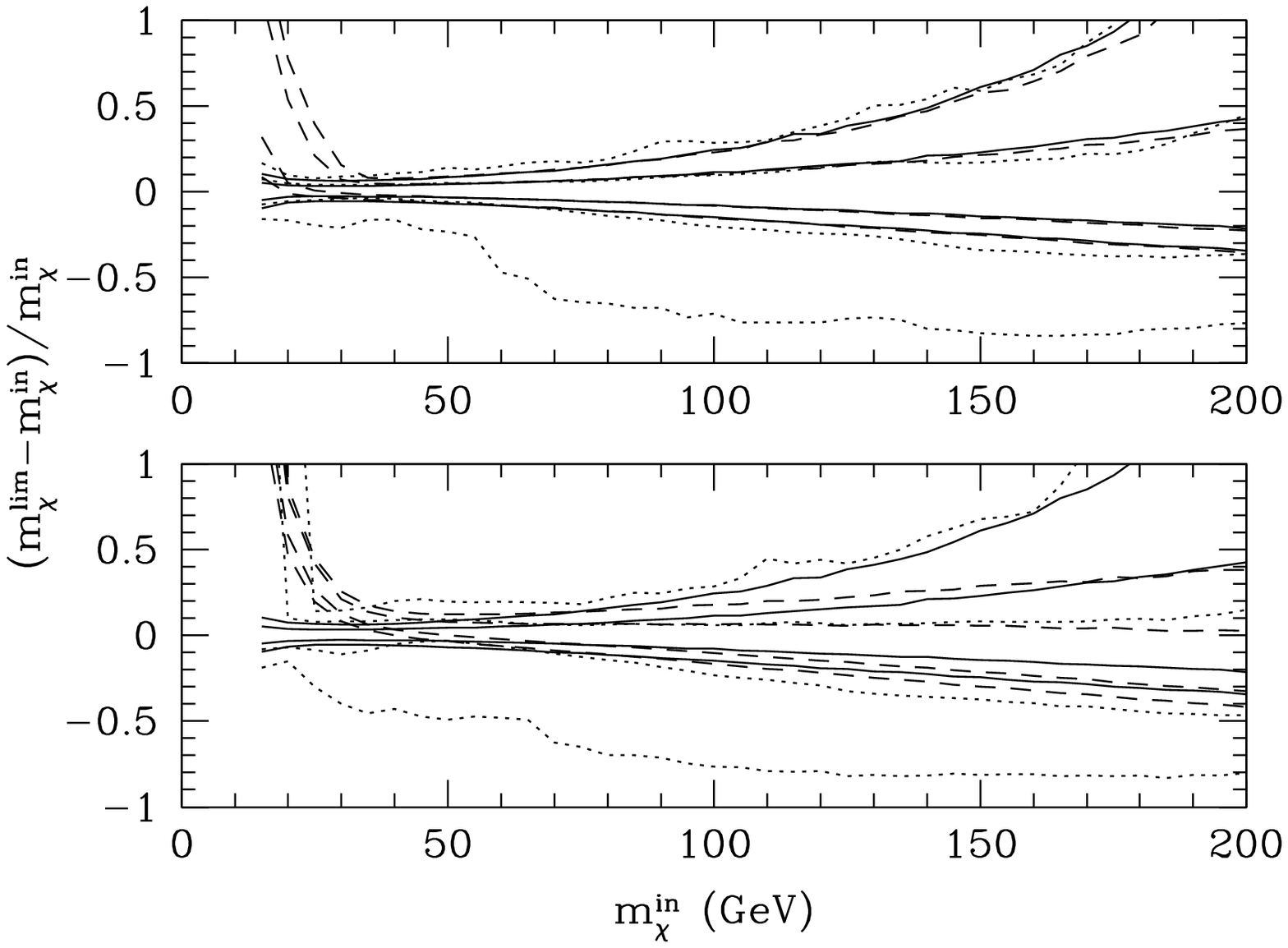} 
\end{center}  
\begin{center} 
  \caption{
As fig.~\ref{back1rat} for an exponential background
energy spectrum with $E_{\rm b} = 20 \, {\rm keV}$.}
\label{back2rat}  
\end{center} 
\end{figure}

The limits on the WIMP mass for an exposure of ${\cal E}= 3 \times
10^{5} \, {\rm kg \, day}$ and $\sigma_{\rm p}=10^{-8} \, {\rm pb}$ in
the presence of a background with an exponential energy spectrum with
$E_{\rm b} = 20 \, {\rm keV}$ are displayed in
fig.~\ref{back2rat}. The exponential background spectrum with $E_{\rm
  b} = 20 \, {\rm keV}$ is similar to the WIMP spectrum with $m_{\chi}
\sim 70 \, {\rm GeV} $. Therefore if the background is neglected in
the likelihood analysis, for smaller (larger) input WIMP masses the
WIMP mass is systematically over(under)-estimated. For the exponential
background when the background rate, $b_{\rm r}$, and characteristic
energy, $E_{\rm b}$, are included in the likelihood analysis there are
still large deviations from the zero background limits, this is
because the WIMP and background spectra can have extremely similar
shapes and can be degenerate. The fluctuations in the limits with
varying $m_{\chi}^{\rm in}$ reflect the errors resulting from this
degeneracy rather than a real underlying trend.

In summary the implications of non-negligible backgrounds for the
determination of the WIMP mass depend strongly on the shape of the
background spectra (as well as, obviously, its amplitude). A flat
background spectrum will lead to a systematic error in the WIMP mass
for light WIMPs, which can be avoided, at the cost of larger
statistical error, by fitting for the background event rate. For
heavier WIMPs the flat background is similar in shape to the WIMP
spectrum and it is hence more difficult to separate the WIMP and
background spectra and accurately measure the WIMP mass. An
exponential background spectrum is similar in shape to the WIMP
spectrum and would inevitably (even when the background parameters are
including in the likelihood analysis) lead to increased errors in the
determination of the WIMP mass.

With a single detector it will be difficult to disentangle a WIMP
signal (and the WIMP mass) from background if the background spectrum
has a similar shape to the WIMP spectrum (i.e. exponential background,
or flat background with a heavy WIMP). Multiple targets (for instance
Ge and Si as used by CDMS II~\cite{CDMS}) would help due to the
dependence of the WIMP spectrum on the mass of the target nuclei.  See
ref.~\cite{material} for Monte Carlo simulations using ${\rm Ca W
  O_{4}}$ and ${\rm Zn W O_{4}}$.  Detectors composed of very
different targets (e.g. Ge and Xe) would likely have different
background spectra however.

\section{Summary}
\label{discuss}

We have studied how the accuracy with which the WIMP mass could be
determined by a single direct detection experiment depends on the detector
configuration and the WIMP properties. Specifically, we investigated
the effects of varying the underlying WIMP mass and cross-section, the
detector target nucleus, exposure, energy threshold and maximum
energy, the local circular speed and the background event rate and
spectrum.

The accuracy of the mass limits is most strongly dependent on the
underlying WIMP mass and the number of events detected. For light
WIMPs (mass significantly less than that of the target nuclei) small
variations in the WIMP mass lead to significant changes in the energy
spectrum.  Conversely for heavy WIMPs the energy spectrum depends only
weakly on the WIMP mass. Consequently it will be far easier to measure
the WIMP mass if it is light than if it is heavy. The number of events
detected is directly proportional to both the exposure and the
cross-section, therefore these quantities have the greatest bearing on
the accuracy of the WIMP mass determination. For our baseline,
SuperCDMS~\cite{SuperCDMS} like, Ge detector with negligible
background and energy threshold $E_{\rm th}= 10 \, {\rm keV}$ for a
WIMP-proton cross-section of $\sigma_{\rm p}= 10^{-8} \, {\rm pb}$, a
factor of a few below the current exclusion limits from the
CDMS II~\cite{CDMS} and Xenon10~\cite{Xenon10} collaborations, with
exposures of ${\cal E}= 3 \times 10^{4}$ and $3 \times 10^{5} \, {\rm
  kg \, day}$ it would be possible to measure the mass of a light
[$m_{\chi} \sim {\cal O}(50 \, {\rm GeV})$] WIMP with an accuracy of
roughly $25\%$ and $10\%$ respectively.  These numbers, and the upper
limits in particular, increase with increasing WIMP mass, and for
heavy WIMPs ($m_{\chi} \gg 100 \, {\rm GeV}$) even with a large
exposure it will only be possible to place a lower limit on the mass.
For very light WIMPs, $m_{\chi} < {\cal O}(20 \, {\rm GeV})$, the
number of events above the detector energy threshold would be too
small to allow the mass to be measured accurately. If $\sigma_{\rm p}=
10^{-9} \, {\rm pb} $, with an exposure of ${\cal E} = 3 \times 10^{5}
\, {\rm kg \, day}$ it will be possible to measure the mass with an
accuracy of $\sim \pm 20 \% \, ({}_{-30\%}^{ + 100\%})$ for
$m_{\chi}^{\rm in} = 50 \, (100) \, {\rm GeV}$. If $\sigma_{\rm p}=
10^{-10} \, {\rm pb} $, for $m_{\chi}^{\rm in} = 50 \, {\rm GeV}$,
even with an exposure of ${\cal E} = 3 \times 10^{5} \, {\rm kg \,
  day}$ it will only be possible to determine the WIMP mass to within
a factor of a few and for more massive WIMPs it will only be possible
to place a lower limit on the mass.

The energy threshold, $E_{\rm th}$, and the maximum energy, $E_{\rm
  max}$, above which recoils are not detected/analysed also affect the
accuracy with which the WIMP mass can be determined. Increasing
$E_{\rm th}$ (or decreasing $E_{\rm max}$) not only reduces the number
of events detected, but also reduces the range of recoil energies and
the accuracy with which the characteristic energy of the energy
spectrum, $E_{\rm R}$, and hence the WIMP mass, can be measured.  The
effect of increasing $E_{\rm th}$ is smallest for intermediate WIMP
masses. For light WIMPs the small $E_{\rm R}$ means that the expected
number of events decreases rapidly as the energy threshold is
increased, while for heavy WIMPs the large $E_{\rm R}$, and flatter
energy spectrum, means that the smaller range of recoil energies
reduces the accuracy with which $E_{\rm R}$ can be measured.  The
effect of reducing the maximum energy (from infinity to $100 \, {\rm
  keV}$) is very small for light WIMPs as the differential event rate
above $E_{\rm max} = 100 \, {\rm keV}$ is negligible, however for
heavy WIMPs the fractional mass limits can change by ${\cal O}(10
\%)$.

The relative capabilities of different detectors to determine the WIMP
mass depend not only on the WIMP and target masses, but also on their
energy thresholds. The WIMP and target mass dependence of the
characteristic energy scale of the recoil spectrum suggests that heavy
targets will be able to measure the mass of a heavy WIMP more
accurately, however the rapid decrease of the nuclear form factor with
increasing momentum transfer which occurs for heavy nuclei means that
this is in fact not the case.

If the WIMP distribution on the ultra-local scales probed by direct
detection experiments is smooth, then the uncertainties in the
detailed shape of the local velocity distribution lead to relatively
small changes in the shape of the differential event
rate~\cite{drdens,lgegreen}, and hence a relatively small, [${\cal O}
(5 \%)$], systematic uncertainty in the WIMP mass~\cite{pap1}.  There
is however an uncertainty in the local circular speed, $v_{\rm c}$,
(and hence the typical speed of the WIMPs) of order $\pm 20 \, {\rm km
  \, s}^{-1}$~\cite{klb} and since $E_{\rm R}$ depends on both
$m_{\chi}$ and $v_{\rm c}$ this leads to a degeneracy between
$m_{\chi}$ and $v_{\rm c}$~\cite{pap1}. For deviations of $\pm 20 \,
{\rm km \, s}^{-1}$ in the underlying value of $v_{\rm c}$ this
systematic error increases with increasing $m_{\chi}^{\rm in}$ from
$\sim 10\% $ for small $m_{\chi}^{\rm in}$ to $\sim 40\%$ for
$m_{\chi} \approx 200 \, {\rm GeV}$.

The assumption of a smooth WIMP distribution may not be valid on the
sub milli-pc scales probed by direct detection experiments (see
discussion in Paper I~\cite{pap1}). If the ultra-local WIMP
distribution consists of a finite number of streams (with a priori
unknown velocities) then the recoil spectrum will consist of a number
of (sloping due to the energy dependence of the form factor)
steps. The positions of the steps will depend on the stream
velocities, the target mass and the WIMP mass. In this case multiple
targets would be needed to extract any information on the WIMP mass.
Drees and Shan~\cite{sd} have recently demonstrated that with multiple
targets it is in principle possible to constrain the WIMP mass without
making any assumptions about the WIMP velocity distribution.

Future experiments aim to have negligible backgrounds, however,
non-negligible neutron backgrounds would lead to errors in the
determination of the WIMP mass. The size of the errors will depend on
the amplitude and shape of the background spectrum.  If the background
rate is not negligible compared with the WIMP event rate it will be
difficult to disentangle a WIMP signal (and the WIMP mass) from the
background if the background spectrum has a similar shape to the WIMP
spectrum (i.e. exponential background, or flat background with a heavy
WIMP). The uncertainties from backgrounds could be mitigated by using
multiple targets (see e.g. Ref.~\cite{material}), however detectors
composed of very different targets (such as Ge and Xe) would be
unlikely to have the same background spectra.

\ack
 AMG is supported by STFC and is grateful to Ben Morgan 
for useful discussions and Marcela Carena for encouragement to investigate
some of the issues considered.

\end{document}